\documentclass[aps,prb,twocolumn]{revtex4}

\usepackage{epsfig}		
\usepackage{dcolumn}		
\usepackage{hyperref}		
\usepackage{bm}

\newcommand\cm{cm$^{-1}$ }		
\newcommand\etal{\textit{et al.}}	
\newcommand\ie{\textit{i.e.}}		

\begin{document}

\title{Intermediate regime in Tetrathiafulvalene-Chloranil (TTF-CA)\\
pressure-induced neutral-ionic transition}

\author{Matteo Masino,  Alberto Girlando}
\affiliation{Dipartimento di Chimica G.I.A.F. and INSTM-UdR Parma,
Universit\`a di Parma, Parco Area delle Scienze, I-43100, Parma, Italy}
\author{Aldo Brillante}
\affiliation{Dipartimento di Chimica Fisica e Inorganica and INSTM-UdR Bologna,
Universit\`a di Bologna, Viale Risorgimento 4, I-40136 Bologna, Italy}

\date{\today}

\begin{abstract}

We report a detailed spectroscopic study of the pressure induced
neutral-ionic phase transition (NIT) of the mixed-stack
charge-transfer (CT) crystal tetrathiafulvalene-chloranil (TTF-CA).
We show that the pressure induced phase transition is still
first-order and involves the presence of an intermediate {\it disordered}
phase, defined by the coexistence of two species of different ionicity.
Further application of pressure gradually converts this phase
into an homogeneous ferroelectric phase with a single ionicity.
In addition, we detect strong pretransitional phenomena which
anticipate the intermediate phase and are indicative of a
precursor dynamic regime dominated by fluctuations.

\end{abstract}

\pacs{Valid PACS appear here}
                             

\maketitle

\section{Introduction} \label{s:Intro}

Neutral-ionic phase transitions (NIT) are pressure or
temperature induced valence instabilities occurring
in mixed stack charge-transfer (CT) crystals, where
$\pi$-electron donor (D) and acceptor (A) molecules alternate
along the stack direction.\cite{girlando04,horiuchi06}
NIT are known since long time,\cite{torrance81,girlando81}
but the microscopic origin of the many intriguing phenomena 
accompanying the transition is still controversial. 
For instance, the dielectric constant and conductivity anomalies 
have been ascribed to the presence of nanoscopic, mobile objects, 
the lattice relaxed neutral-ionic domain walls (NIDW), also referred 
as lattice relaxed charge-transfer exciton strings (LRCT).\cite{nagaosa86,lemee97,luty02} 
On the other hand, the dielectric constant anomaly can be well 
accounted for by the charge oscillations induced by the 
Peierls mode,\cite{freo02,soos04} since the ionic stack
is subject to Peierls instability.\cite{masino06}  
Another intriguing aspect of NIT is indeed associated
with the dimerization Peierls instability, since dimerized ionic
stacks are potentially ferroelectric.

Tetrathiafulvalene-Chloranil (TTF-CA) NIT is by far the
most widely studied, also because it can be induced both
by temperature $T$ and pressure $p$. 
The ambient pressure $T$-induced transition occurs at $T_c$= 81 K,
is first order, and has been well characterized by a series of 
structural and spectroscopic studies, whose list is too long 
to be reported here.\cite{horiuchi06,girlando04}

At $T_c$, the average charge on the molecular sites (ionicity, $\rho$) 
jumps from about 0.3 to over 0.5, the latter value being considered 
the borderline between neutral ({\it N}) and ionic ({\it I}) phases. 
The ionicity jump is accompanied by  the stack dimerization.
The Peierls modes have been identified in both
the {\it N} and {\it I} phases,\cite{masino03,masino06}
but of course they do not reach zero frequency, as the
first order valence instability takes over.

The TTF-CA $p$-induced transition has been
comparatively less studied. Early infrared (IR)
spectra of the powders showed that above $\sim$ 1.1 GPa
the phase is ionic and dimerized, similarly
to the low-$T$ phase.\cite{tokura86,girlando86}
On the other hand, preliminary X-ray measurements,
never published in complete form, suggested
that the $p$-induced transition is different
from the $T$-induced one, occurring through some sort
of ``intermediate transition region'', from 
about 0.65 to 1.1 GPa.\cite{metzger85} This finding
has been confirmed by several spectroscopic
data,\cite{tokura86,kaneko87,takaoka87,mitani87,hanfland88,okamoto89}
but without agreement about the nature of this intermediate
transition region. Several authors suggested that species
of different ionicity were present, but the degree of
ionicity was highly uncertain. 
Also the pressure interval of the coexistence region varied depending on 
the type of measurement.\cite{tokura86,kaneko87,okamoto89}
The use of powdered samples is certainly one of the reasons of such 
uncertainties, as for instance IR spectra are more difficult to
interpret in the lack of polarization information. 
Moreover, it is known that the NIT is affected by the presence of
defects, and these certainly dominate in powdered samples. 
In any case, qualitative phase diagrams started to emerge,
\cite{mitani87,takaoka87} based on measurements in which 
both $T$ and $p$ were simultaneously changed. 
According to these studies, both neutral and ionic species
coexist in the intermediate regime.

The first detailed study of the phase diagram, performed
through neutron diffraction and NQR on TTF-CA single
crystals,\cite{lemee97,luty02} definitely
evidenced a bifurcation of the \textit{N-I} crossing line at
a triple point located around 210 K and 0.5 GPa.
According to these authors, the intermediate regime arising
above the triple point has to be interpreted as a paraelectric
phase, separating the ionic and dimerized (ferroelectric)
from the neutral and regular stack phase. 
In other words, the intermediate regime is interpreted 
as a true thermodynamic phase, where dynamically 
disordered LRCT are present. 
The condensation and ordering of LRCT, driven by 
interstack interactions, yields the ferroelectric phase.

Evidence of LRCT was thought to be provided
by IR spectra.\cite{okamoto89,lemee97}
However, by using combined single crystal IR and Raman
data, some of us have shown that the apparent signatures
of LRCT in the $T$-induced NIT are actually due to the
Peierls mode.\cite{masino03} For this reason we have
decided to re-investigate the $p$-induced transition of
TTF-CA by the same methods, namely, combined single crystal
Raman and polarized IR data. A first set of experiments,
focalized on the IR spectra polarized perpendicular to the
stack, have clearly shown that two species of different ionicity
are present in the intermediate transition region.\cite{masino04}
In pressure experiments particular care has to be paid to pressure
homogeneity on the sample,  
and before publishing a full paper we have
carefully repeated the experiment, adding also
measurements with polarization parallel to the stack.
In the meantime, an independent paper has appeared,
reporting single crystal optical spectra
as a function of pressure.\cite{matsuzaki05} The 
reported IR spectra confirm our earlier experiment,
but the interpretation is quite different from ours.

In the present paper we report the complete IR absorption spectra
of TTF-CA single crystal, together with  Raman spectra 
collected at the same pressures. 
We confirm that between 0.86 and 1.24 GPa species
with different ionicity are simultaneously present.
However, both species are on the \textit{I} side. In addition,
a precursor regime, probably dominated by fluctuations,
is present between $\sim$ 0.6 and $\sim$ 0.86 GPa.
Possible scenarios emerging from the present measurements
are discussed.








\section{Experimental} \label{s:Exp}

TTF-CA has been prepared by mixing hot saturated
acetonitrile solutions of commercial grade TTF and CA.
Very thin single crystals suitable for infrared (IR) absorption
have been obtained by subliming TTF-CA under reduced pressure.\cite{masino03} 
Thicker samples have been used for the Raman measurements.  
The crystals present significant dichroism in white light transmission:
they appear green when light is polarized parallel to the stack axis
{\it a}, and yellow for light polarized perpendicular to the stack.

The IR absorption spectra have been measured with a Bruker FTIR
spectrometer (model IFS66), equipped with a microscope. The Raman
spectra were recorded with a Renishaw System 1000 micro-spectrometer
(20X magnification objective) and using the 568.2 nm excitation line 
from a Kr ion laser.
The laser power has been adjusted to 1 mW in order to reduce the risks 
of sample heating especially in proximity of the phase transition.
The spectral resolution of both IR and Raman spectra is 2 \cm.

High pressure measurements up to 3.2 GPa have been performed with 
a custom designed diamond anvil cell (DAC) able to fit under both 
IR and Raman microscopes. The samples have been inserted in either  
stainless-steel or copper gaskets, the latter
allowing finer tuning at low to 
moderate pressures (up to 1.5 GPa).  
Pressure calibration has been done with the ruby luminescence
technique.\cite{ruby}
Estimated error bar in the pressure reading is $\pm 0.05$ GPa.
Liquid paraffin (Nujol oil) has been used as the pressure
transmitting medium both in IR and Raman.
In IR, the spectral regions between 1360-1390 \cm 
and 1430-1480 \cm are obscured by the Nujol bands.

\section{Results}
\label{s:Res}

\subsection{Valence instability: Ionicity}
\label{ss:rho}

\begin{figure}[h]
\includegraphics[width=7.8cm]{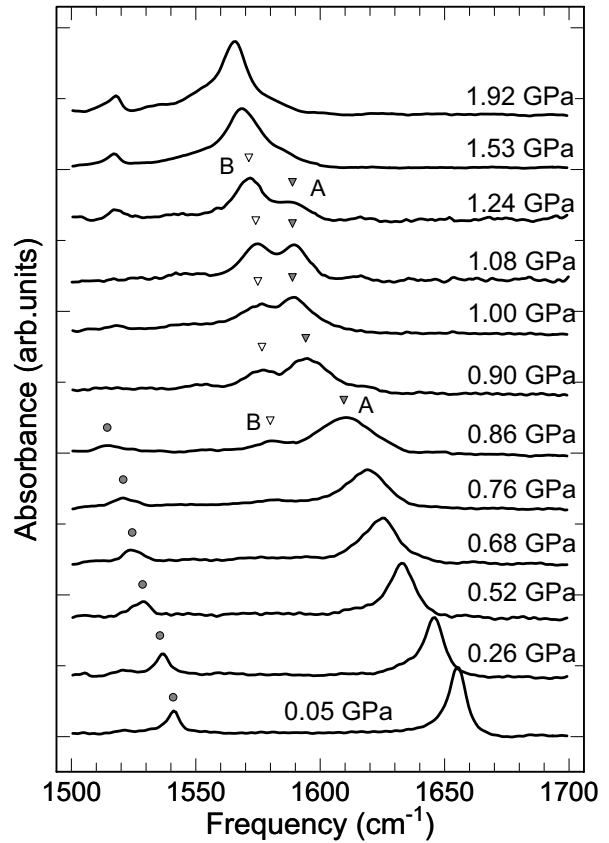}
\caption{Polarized IR absorption spectra of TTF-CA as a
function of pressure. Light polarized perpendicular to the stack 
direction ({\bf E} $\perp a$).
Open and solid triangles mark the the two bands assigned
to the $b_{1u}$ C=O stretching mode in the coexistence regime. 
Solid circles indicate the band due to the
$b_{2u}$ C=C stretching mode of CA molecules.} 
\label{f:fig1}
\end{figure}

In Fig.\ref{f:fig1} we show the IR absorption spectra polarized
perpendicular to the stack axis in the 1500-1700 \cm 
spectral region as a function of pressure.
Here two bands at 1655 and at 1542 \cm (spectrum at 0.05 GPa) are
clearly observed, and can be safely assigned to the $b_{1u}$ C=O
and the $b_{2u}$ C=C stretching modes of the CA molecular components, 
respectively. \cite{girlando83}
Both C=O and C=C bonds are strongly affected by the $\pi$-electronic structure
of the molecule and lose part of their bond order when an electron is
added. For this reason both vibrational modes are sensitive to 
the effective charge localized on the molecules, \ie,~ the ionicity. 
In particular the $b_{1u}$ C=O band is very useful to study the NIT
valence instability, and can 
be used to accurately probe ionicity in view of its large 
frequency shift in going from the fully neutral CA$^0$ ($\omega=1685$ \cm) 
to the fully ionic CA$^-$ molecule ($\omega=1525$ \cm).\cite{girlando83}

From Fig.\ref{f:fig1} it is evident that the $b_{1u}$ C=O band displays
an anomalous behavior with increasing pressure.
After a smooth softening accompanied by a sizeable broadening,
it splits giving rise to two bands, marked A and B in the figure.
The B band starts to appear around 0.86 GPa, at 1580 \cm,
on the low frequency side of the A band at 1610 \cm. With 
increasing pressure the B band develops, gaining intensity at 
the expenses of the A band. Then above 1.24 GPa the intensity
inversion is completed, the A band disappears, while the B band 
is observed at 1570 \cm and remains nearly unchanged 
up to the maximum experimental pressure we have reached (3.2 GPa). 

\begin{figure}[h]
\includegraphics[width=7.8cm]{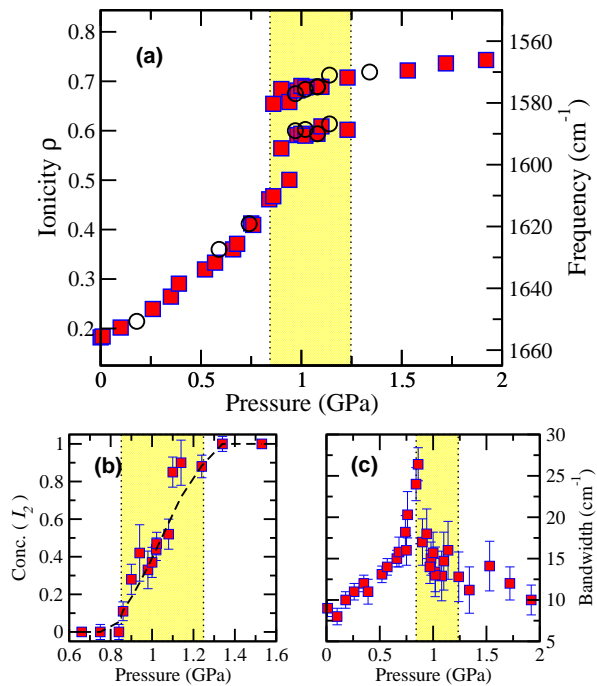}
\caption{(color online). (a): Pressure evolution of TTF-CA ionicity, $\rho$,
as estimated by the frequency of the $b_{1u}$ C=O antisymmetric
stretching mode. The yellow region marks the pressure range of the
coexistence regime. (b): Relative concentration of the $I_2$ species
as estimated from the intensity ratio of the ``B band'' component.
The dashed line is a guide for the eye.
(c): Bandwidth (FWHM) of the $b_{1u}$ C=O band as a function of pressure.}
\label{f:fig2}
\end{figure}

Ionicity values deduced from the experimental frequencies 
of the $b_{1u}$ C=O band by assuming a linear frequency 
dependence are plotted in Fig.~\ref{f:fig2}a.
Ionicity increases gradually from $\rho$ = 0.19 up to 
$\rho$ = 0.47 at 0.86 GPa. 
Applying pressure further a small discontinuity occurs.
The TTF-CA system crosses the \textit{N-I} borderline,
conventionally located at $\rho$ = 0.5.
An intermediate regime develops, where molecular 
species with two different ionicity, $\rho(I_1)$ and 
$\rho(I_2)$, coexist.
$\rho(I_1)$ slightly increases from $\sim 0.55$ up to $\sim 0.60$,
while $\rho(I_2) \sim 0.7$ is almost constant.
These two different ionicities correspond to the two components,
A and B, of the splitted band structure in Fig.\ref{f:fig1}, 
and the evolution of their relative intensities allows one 
to follow the concentrations, $c(I_1)$ and $c(I_2)$, 
of the two molecular species ($c(I_1) + c(I_2) = 1$).
The two species have variable concentration as demonstrated
by Fig \ref{f:fig2}(b), where $c(I_2)$, estimated from the relative 
intensity of the B band, is plotted as a function of pressure.
The concentration $c(I_2)$ increases from 0.86 up to 1.24 GPa, 
and above this pressure only $I_2$ molecules 
with $\rho \sim 0.7$ are present.

In Fig.~\ref{f:fig2}c we report the pressure evolution 
of the bandwidth of the $b_{1u}$ C=O band.
Here the discontinuity between 0.86 and 0.90 GPa 
in crossing the \textit{N-I} borderline,
is much clearer. Also notice the steep increase of bandwidth
when approaching the coexistence region, indicative of 
a precursor regime with strong dynamic disorder. 

We remark that the pressure dependence of the $b_{2u}$ C=C band at
1542 \cm, marked with solid circles in Fig.~\ref{f:fig1}, 
is perfectly consistent with that of the $b_{1u}$ C=O band. 
The frequency softening, followed by a gradual loss of intensity,
indicates that ionicity smoothly increases up to 0.86 GPa.
Above this pressure, when the system has shifted to the 
ionic side ($\rho > 0.5$), this band cannot be detected 
anymore in our spectra. 
This behavior is a consequence of the fact that
in CA$^-$  the description of this vibrational mode
in terms of C=C stretching changes drastically, and its
IR intensity falls down almost completely.\cite{ranzieri07}

The above arguments validate our analysis of the 
1500 - 1700 \cm spectral region and rule out the alternative 
interpretation proposed by Matsuzaki \etal ,\cite{matsuzaki05}
based on the incorrect assignment of the B band in Fig.~\ref{f:fig1} 
to the $b_{2u}$ C=C stretching band of the CA molecular units.

Finally we notice that the occurrence of the coexistence phase
has been carefully checked by repeated pressure cycling on different 
TTF-CA samples. We can also rule out pressure inhomogeneity across
the sample, since we have checked that Raman micro-spectroscopy, 
with a spatial resolution of about 5 $\mu$m, gives identical 
spectra throughout the whole sample. 
Moreover open circles in Fig.~\ref{f:fig2}a, 
representing experimental points taken on releasing pressure from 3.2 GPa,
indicate that this phase is fully reversible and that hysteresis 
effects have not been detected.

\subsection{Structural instability: Dimerization}
\label{ss:delta}

At room temperature and at ambient pressure 
TTF-CA crystallizes in the $P2_1/n$ space group
with two formula units per cell.\cite{lecointe95} 
The TTF and CA molecules alternate with
uniform distance along the crystallographic $a$ axis,
each molecule residing on inversion symmetry center.
The temperature induced NIT is accompanied
by a dimerization instability, and at low temperature 
the space group is $P_n$, with two dimerized stack 
per unit cell.\cite{lecointe95}

It is well known that IR spectroscopy with light polarized parallel
to the stack direction is a powerful method to study the
dimerization instabilities in CT crystals.\cite{girlando83}
The loss of the inversion symmetry associated with the 
stack dimerization makes the Raman active totally-symmetric 
molecular vibrations also IR active.
The totally-symmetric modes are coupled to CT electrons
(\textit{e-mv} coupling), and in IR they borrow huge intensity 
from the nearby electronic CT transition, with polarization
parallel to the stacks.
One can therefore discriminate between dimerized and uniform 
stack structure on the basis of the presence or absence of 
these characteristic IR bands (``vibronic bands'').\cite{girlando83} 

\begin{figure}[h]
\includegraphics[width=7.8cm]{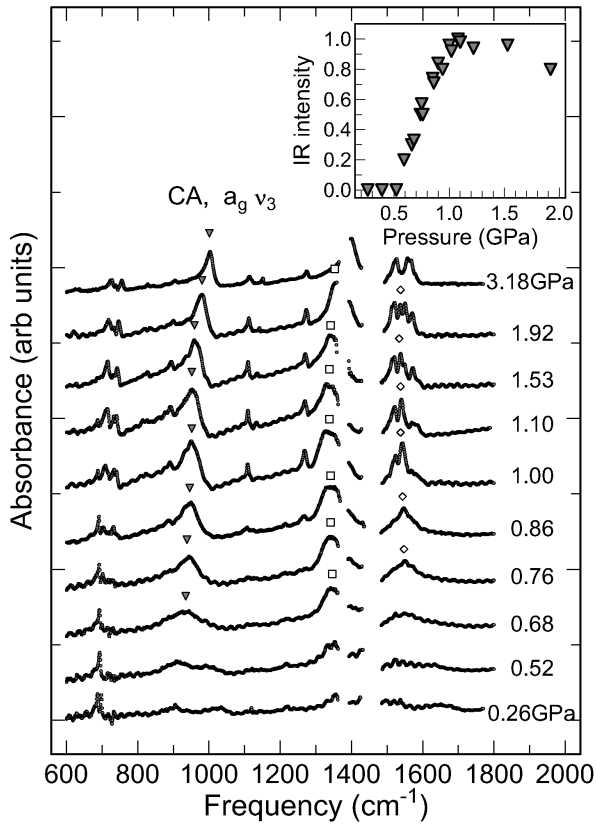}
\caption{Polarized IR absorption spectra of TTF-CA as a
function of pressure. Light polarized parallel to the stack 
direction ({\bf E} $\parallel a$). The spectral regions
between 1360-1390 \cm and 1430-1480 \cm is not accessible
due to absorption of the pressure medium.
Inset: Normalized IR intensity of the CA $a_g$  $\nu_3$ band.}
\label{f:fig3} 
\end{figure}

In addition, it has been recently demonstrated that 
IR spectra parallel to the stack also yield
useful information on pretransitional dynamics
driving the dimerization instability.
TTF-CA spectra in the neutral, uniform
stack phase are indeed characterized by the
presence of two-phonon excitations (IR ``side-bands''),
whose temperature evolution probes the energy of the
soft mode involved in the stack distortion.\cite{masino03}

To investigate these structural and pretransitional phenomena
of pressure induced NIT of TTF-CA, we have collected 
IR spectra polarized parallel to the stack axis
as a function of pressure (Fig.~\ref{f:fig3}).
Three strong bands, marked by a triangle,
square and circle in the figure, clearly develop
above $\sim$ 0.6 GPa.
These vibronic bands are assigned to the most strongly 
\textit{e-mv} coupled totally-symmetric modes,
namely the CA~ $a_g$ $\nu_3$ around 980 \cm, 
and the TTF $a_g$ $\nu_3$ and $\nu_2$ modes 
around 1350 and 1540 \cm, respectively.
Their presence  reflects the fact
that dimerization is taking place on increasing pressure.\cite{girlando86}

In the inset of Fig.~\ref{f:fig3} 
we report the pressure dependence of the normalized 
IR intensity of the CA $a_g$ $\nu_3$ mode. 
This mode starts to appear as a broad weak band
above 0.6 GPa and rapidly gain intensity
reaching its maximum value around 1.0 GPa, 
A slight intensity decrease follows on increasing pressure further.
The behavior of the other vibronic bands is 
qualitatively similar to this one, although it is not possible 
to carefully follow their intensity evolution, 
because the TTF $a_g$ $\nu_3$ band is partially 
covered by absorption of the pressure transmitting medium, 
while the $a_g$ $\nu_2$ band at 1540 \cm occurs in a region 
overlapped by other fundamental vibrational bands.

\begin{figure}[h]
\includegraphics[width=7.8 cm]{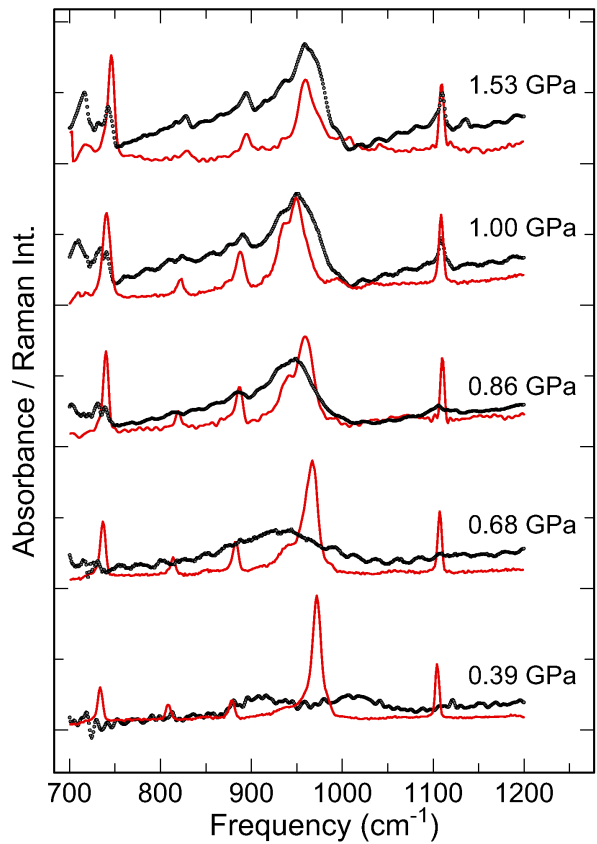}
\caption{(color online). Combined IR and Raman spectra of TTF-CA 
as a function of pressure in the spectral region of the CA
$a_g$  $\nu_3$ vibronic band. Black dots: IR spectra polarized 
parallel to the stack direction ({\bf E} $\parallel a$). 
Red line: unpolarized Raman spectra.}
\label{f:fig4}
\end{figure}

The comparison between the inset of Fig.~\ref{f:fig3} with
Fig.~\ref{f:fig2}a seems to indicate that the dimerization 
and valence instability are not correlated in the pressure 
induced NIT of TTF-CA. 
The onset of dimerization appears to be around 0.6 GPa,
reaching saturation at 1.0 GPa, whereas the double ionicity 
region starts at 0.86 and extends to about 1.24 GPa.
However, in addition to the IR intensity of the
CA $a_g$ $\nu_3$ mode, one has to consider its bandshape,
and make a proper comparison with the corresponding
Raman band. To such aim, we report in Fig.~\ref{f:fig4}
the pressure evolution of an enlarged portion
of the IR spectrum polarized parallel to the stack,
together with Raman spectra collected at the same pressures.


We analyze Fig.~\ref{f:fig4} starting from low pressures. 
Below 0.68 GPa  we can easily identify the so-called side-bands,
namely IR bands occurring above and below
the $a_g$  $\nu_3$ Raman band. The side-bands
are clearly due to sum and difference two-phonon 
excitations between the Raman active molecular vibration 
and a low frequency lattice mode also coupled
with the CT electrons. In the temperature
induced transition of TTF-CA, the side-bands provide useful 
information on the soft mode which drives the dimerization 
instability, namely, the Peierls mode.\cite{masino03} 
At ambient conditions the side-bands are separated
by about 70 \cm from the central Raman band.
At 0.39 GPa, the distance is about 50 \cm,
approximately as in the 150 K spectrum at ambient pressure.
In our experiments IR side bands can be clearly identified up 
to about 0.52 GPa, and disappear, merging
in an unique broad band on further increasing pressure.

Indeed, above $\sim$ 0.6 we observe the development
of a single vibronic IR band due to 
the CA $a_g$ $\nu_3$. This band is initially very
broad (almost 100 \cm), with the maximum occurring
at a frequency \textit{lower} than the corresponding Raman
band. As the pressure increases, the vibronic band
sharpens and its frequency moves upwards, approaching 
the value of the Raman band. Around $\sim$ 0.9 GPa 
(but it is difficult to set a precise pressure, as
we are comparing IR and Raman bandshapes) we start to
see coincidence between the IR vibronic and the Raman
band, perfect matching occurring at 1.0 GPa and above.  
The lack of IR and Raman frequency
coincidence proves that the loss of the inversion 
symmetry is not complete along the stack,
although some kind of local stack distortion has already taken place.
This phenomenon is indicative of a new regime
established between  $\sim$ 0.6 GPa and $\sim$ 0.9 GPa 
and dominated by dynamic disorder.
Notice that in this pressure range also the IR spectra
polarized perpendicular to the stack show evidence
of strong dynamic disorder (Fig.~\ref{f:fig2}c).

The IR spectra polarized perpendicularly to the
stack show that above $\sim$ 0.86 GPa we enter
in a regime characterized by a double ionicity.
This pressure is approximately the same in which
we start to see coincidence between Raman
and IR vibronic bands, both of which display
a clear doublet structure. This regime is then
characterized by comparable domains of
different ionicity, and around 1.0 GPa we
can affirm that the stacks are in any case fully dimerized,
since the intensity of the vibronic bands
saturates (inset of Fig.~\ref{f:fig3}).
As we have seen in the previous section, the
double ionicity regime gradually evolves
towards a single ionicity, which is finally
reached above $\sim$ 1.24 GPa.




\section{Discussion and Conclusions}\label{s:conclusions}

As already stated, TTF-CA $T$-induced NIT at ambient pressure
is well characterized. It is a first order 
transition with a discontinuous $\rho$ jump from  $\sim$ 0.3
to just over 0.5, and a simultaneous dimerization of the stacks.
The dimerization is anticipated by a critical softening of 
an effective Peierls mode which proves the displacive nature
of the structural phase transition.\cite{masino03}
On the other hand, the evolution of TTF-CA $p$-induced NIT is 
more complex, actually more than it has been thought so far.

The $p$-induced NIT is less discontinuous than the $T$-induced one,
but is still first order since a weak discontinuity in the $\rho(p)$
curve is observed between 0.86 and 0.9 GPa (Fig.~\ref{f:fig2}a).
Actually, if we follow the pressure evolution of the ionicity $\rho$,
three different regimes are identified.
A neutral phase ($\rho < 0.5$) at low pressure,
a coexistence phase between 0.86 and 1.24 GPa, 
characterized by species of two different ionicities, 
and a ionic phase ($\rho$ $\sim$ 0.7) at high pressures.

In the neutral phase the ionicity grows continuously with pressure,
and reaches $\rho$ = 0.47 at 0.86 GPa, the transition pressure.
However, strong pretransitional phenomena related to both the 
valence and the structural instability start already 
above $\sim$ 0.6 GPa, when the ionicity is about 0.3.
First, the bandwidth of the modes most sensitive to
$\rho$ shows a sharp increase with $p$, 
starting above 0.6 GPa and reaching the maximum at the 
critical pressure $p_c$ = 0.86 GPa (Fig.~\ref{f:fig2}c).
Second, approximately in the same pressure interval, 
the IR spectra polarized parallel to the stack are characterized 
by vibronic bands, corresponding to the activation in IR of 
intramolecular totally-symmetric modes. 
The activation, with appreciable intensity due to the
coupling to the CT electrons, is the signature of a dimerization distortion.
However, completely dimerized stacks would lead to IR-Raman 
frequency coincidence, due to the loss of the inversion 
symmetry along the stacks.
IR-Raman frequency coincidence is \textit{not} observed
in the pressure interval we are considering 
(Figs.~\ref{f:fig3} and ~\ref{f:fig4}). 
This finding can be rationalized in terms of localized distortions,
most likely fluctuating in time and space, on the average preserving 
the inversion symmetry.
In other words, the $p$-induced phase transition at room $T$ 
is preceded by a {\it precursor regime} characterized by fluctuating
dimerized domains with variable $\rho$ in the predominantly
neutral and regular stack phase. 

The fluctuating behavior fades around $p_c$ = 0.86 GPa,
as evidenced by Fig.~\ref{f:fig2}c, and TTF-CA develops 
an intermediate phase which persists up to 1.24 GPa.
In this pressure range the $b_{1u}$ C=O band, which probes the ionicity, 
shows a well resolved doublet structure (Fig.~\ref{f:fig1}), 
the IR intensity of the vibronic bands reaches the maximum value
and frequency matching between IR and Raman bands
is definitely established (Figs.~\ref{f:fig3} and ~\ref{f:fig4}).
These experimental findings indicate that the intermediate phase 
is characterized from the structural point of view by the lack of 
inversion symmetry, and from the electronic point of view by
the coexistence of species with different ionicity, 
$\rho(I_1) \sim 0.6$ and $\rho(I_2) \sim 0.7$.
The intensity evolution of the C=O doublet shows that the $\rho(I_1)$ 
component gradually converts into the more ionic one, 
$\rho(I_2)$ (Fig.~\ref{f:fig2}b).
We then argue that this phase is not a well defined, 
ordered phase with a spatially modulated ionicity parameter, 
but is properly described in terms of a static disordered phase 
where species with two different ionicity coexist.

On increasing pressure further, coexistence is smoothly 
suppressed and, above 1.24 GPa, TTF-CA is ionic ($\rho \sim 0.7$) 
and dimerized similarly to the low-$T$ ferroelectric phase.
We do not detect discontinuities or significant spectral changes, besides the vanishing 
of the $\rho(I_1)$ species in favor of the more ionic $\rho(I_2)$.
Accordingly, we suggest that the phase transformation
from the intermediate to the final ionic phase occurs continuously
through a disorder-order type mechanism.
\vskip 0.5cm

\begin{figure}[h]
\includegraphics[width=7.8 cm]{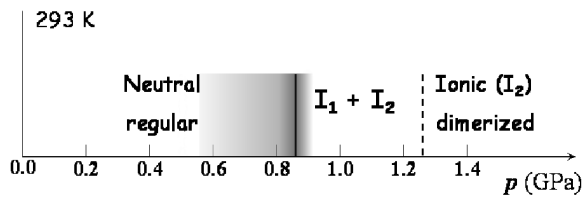}
\caption{TTF-CA phase diagram at $T = 293 K$, variable $p$.}
\label{f:fig5}
\end{figure}

Fig.~\ref{f:fig5} reports a sketch of the just described
TTF-CA phase evolution on increasing $p$ at ambient
temperature. Several hypotheses have been 
formulated about the nature of the 
intermediate regime between the neutral,
regular and the ionic dimerized stack phase.
Up to now, the most credited idea has been that of a ``paraelectric''
phase, dominated by fluctuating, disordered LRCT,
which precede and induce the tridimensional ordering
to the ferroelectric (ionic and dimerized) phase.\cite{lemee97,luty02}
Our results suggest that we have a precursor regime,
with fluctuations (shaded area in Fig.~\ref{f:fig5}),
\textit{and} a new phase, characterized by the simultaneous
presence of two different ionicities, which precedes
the ferroelectric phase. 

We believe that thermally accessible low-lying excitations 
such as LRCT \cite{luty02, collet02} are present in the precursor regime
between $\sim$ 0.6 and $\sim$ 0.86 GPa. 
Other experimental data are in support of this idea. 
Single crystal reflectivity data show coexistence of neutral
and ionic species between $\sim$ 0.3 - 0.4 GPa 
and $\sim$ 0.9 - 1.0 GPa.\cite{kaneko87,matsuzaki05}.
In the electronic spectra the boundaries of the precursor
regime appear slightly wider than those reported in Fig.~\ref{f:fig5}.
On the other hand, such boundaries cannot be sharp,
and electronic data are more sensitive than vibrational ones.
Dielectric response as a function of pressure
shows anomalies in the relaxation frequency, starting at about
0.5 GPa (the data extend only to 0.7 GPa).\cite{okamoto91}
Furthermore, the dc conductivity increases exponentially with $p$,
the maximum value being reached at about 0.87 GPa,\cite{mitani87}
the critical pressure for the insurgence of the $I_1 + I_2$ phase.
All these data have been explained in terms of the presence of 
LRCT,\cite{kaneko87,matsuzaki05,okamoto91,mitani87} and our data 
in the precursor regime can be interpreted in the same way. 

At this point, it is interesting to compare the
precursor regime of the $p$-induced NIT with that
of the $T$-induced NIT. LRCT have been invoked in both
cases,\cite{matsuzaki05,okamoto91,mitani87}
but actually the experimental results are different,
and can be interpreted in a different way.
For instance, by lowering the temperature the
dielectric constant shows a dramatic increase
as $T_c$ = 81 K is approached, whereas the $p$
variation is less remarkable, and evidenced by
looking at the relaxation frequency change.\cite{okamoto91}
The $T$-dependence of the dielectric anomaly can
be \textit{quantitatively} explained as due to the
charge oscillations induced by the Peierls soft mode,\cite{freo02,soos04} 
and evidence of the soft mode has been achieved by looking at
the IR side-bands\cite{masino03} formerly considered 
as experimental proofs of LRCT.\cite{lemee97}
Finally, if the conductivity increase in the
$p$-induced NIT is due to current carrying LRCT,\cite{okamoto91}
their presence and role is only marginal in driving 
the $T$-induced NIT, as the conductivity shows an increase 
only within a few degrees before $T_c$.\cite{mitani87} 
Recent model calculations have indeed shown
that LRCT are thermally accessible at ambient
temperature, whereas their concentration is very
small at $T_c$.\cite{soos07} 
Metastable domains are then important in TTF-CA 
$p$-induced NIT, but not in the $T$-induced one.

According to the above description, LRCT characterize
the precursor regime of TTF-CA $p$-induced NIT.
However, they do not anticipate and build the ionic, 
ferroelectric phase, but a double ionicity phase, $I_1 + I_2$, 
whose nature we shall now address. 
First of all, it is not the paralectric phase described 
in Ref.~\onlinecite{lemee97}, since domains of different 
ionicity have comparable concentrations, and well 
defined ionicity (Figs.~\ref{f:fig1} and ~\ref{f:fig2}a).
It is therefore difficult to imagine that the
phase between 0.86 and 1.24 GPa is a dynamic phase, 
characterized by mobile nanoscopic domains of different nature. 
Also, a mixed phase containing both {\it N} and {\it I}
stacks has been proposed several times, both
from the standpoint of theory,\cite{hubbard81}
and of experiment.\cite{takaoka87,hanfland88} 
However, our data strongly suggest that the two species 
with different ionicity are both on the ionic side, 
$\rho(I_1) \sim 0.6$ and $\rho(I_2) \sim 0.7$.
We have several hints in this direction. 
We remind only that at 1.0 GPa the intensity of the 
vibronic bands saturates (inset of Fig.~\ref{f:fig3}), 
indicating completely dimerized stacks, with ionicity values 
well above the conventional $\rho$ = 0.5 borderline.\cite{nota}
Thus the intermediate phase is likely a \textit{statically disordered}
mixed phase, defined by randomly oriented \textit{polar}
domains of different ionicity, which gradually polarize
evolving towards the ordered ferroelectric phase above 1.24 GPa.

The characterization of the precursor regime and 
the found coexistence of two different \textit{ionic} domains 
in the intermediate phase are the most novel results 
of this work. From this perspective, we may also envision a 
mechanism of TTF-CA $p$-induced NIT involving a competition 
between nucleation and a sort of solid state spinodal decomposition.\cite{chaikin95}
In this picture the precursor regime is the nucleation regime, 
characterized by large amplitude local fluctuations (droplets). 
Rather than evolving in the new stable phase when the droplets 
size reaches a critical value, we may have the insurgence of small 
amplitude large scale fluctuations of the order parameter, 
so that the system is globally unstable, and immediately 
develops the mixed $I_1 + I_2$ phase, with strongly 
intermixed mesoscopic domains.
Further application of pressure converts this phase into
the ferroelectric phase with a single ionicity.
However, additional measurements, and in particular
high resolution structural analysis, are needed before 
we can give credit to the just sketched appealing 
but highly speculative scenario.

\section{Acknowledgments}
Work supported by the ``Ministero dell' Universit\`a
e Ricerca'' (MUR), through FIRB-RBNE01P4JF and PRIN2004033197\_002.
Many useful discussions with A. Painelli and Z. G. Soos are gratefully
acknowledged.

\end{document}